# An Assessment of Data Transfer Performance for Large-Scale Climate Data Analysis and Recommendations for the Data Infrastructure for CMIP6

Eli Dart, Michael F. Wehner, Prabhat
Lawrence Berkeley National Laboratory, 1 Cyclotron Road, Berkeley, California
94720 USA

**Abstract**
We document the data transfer workflow, data transfer performance, and other aspects of staging approximately 56 terabytes of climate model output data from the distributed Coupled Model Intercomparison Project (CMIP5) archive to the National Energy Research Supercomputing Center (NERSC) at the Lawrence Berkeley National Laboratory required for tracking and characterizing extratropical storms, a phenomena of importance in the mid-latitudes.  We present this analysis to illustrate the current challenges in assembling multi-model data sets at major computing facilities for large-scale studies of CMIP5 data. The characterization of extratropical storms enabled by this work is an example of an entire class of studies that require the assembly (and therefore the transfer) of large volumes of data from the globally-distributed CMIP5 archive to a computing center with sufficient computing power to run the analysis. Because of the larger archive size of the upcoming CMIP6 phase of model intercomparison, we expect such data transfers to become of increasing importance, and perhaps of routine necessity. We believe it is critical that a concerted effort be made to improve data transfer performance between the Earth System Grid Federation (ESGF) data centers which host the CMIP5 archive and major computing facilities which are capable of running large-scale analyses of CMIP5 (and, soon, CMIP6) data.

We find that data transfer rates using the ESGF are often slower than what is typically available to US residences (which is far below what should be the norm for major research institutions) and that there is significant room for improvement in the data transfer capabilities of the ESGF portal and data centers both in terms of workflow mechanics and in data transfer performance.  We believe performance improvements of at least an order of magnitude are within technical reach using current best practices, as illustrated by the performance we achieved in transferring the complete raw data set between two high performance computing facilities (the ALCF and Argonne National Laboratory and NERSC and Lawrence Berkeley National Laboratory). To achieve these performance improvements, we recommend: that current best practices (such as the Science DMZ model) be applied to the data servers and networks at ESGF data centers; that sufficient financial and human resources be devoted at the ESGF data centers for systems and network engineering tasks to support high performance data movement; and that performance metrics for data transfer between ESGF data centers and major computing facilities used for climate data analysis be established, regularly tested, and published.

# 1. Introduction

The free distribution of climate model output data has inarguably increased our knowledge of the climate system and its sensitivity to human induced changes by enabling an extremely diverse scientific community to creatively analyze the output of these computationally intensive simulations. As climate models have increased in sophistication, both the demand for their results and the difficulties in distributing the data have increased. For instance, characterizing weather extremes and the changes in frequency and intensity of extreme events under global warming is an important research problem often requiring daily averaged model output (Sillman et. al. 2013; Kharin et. al. 2013). Furthermore, as climate models continue towards finer spatial resolution, the simulation of individual storms and their statistics becomes more realistic (for instance, Wehner et. al. 2014). Hence, sub-daily model output to track these individual simulated storms has become more interesting, furthering the demand for large data transfers from the model repositories to major computing facilities for analysis by individual researchers.

Large multi-model climate data archives such as the Coupled Model Intercomparison Project Phase 5 (CMIP5) archive (Taylor et. al. 2012) present an invaluable resource to understand the effect of climate change on these and other weather phenomena by sampling several sources of projection uncertainty, including structural, internal and forcing uncertainties. However, processing multi-terabyte datasets such as those in the CMIP5 archive is a daunting task for most climate data analysts., with the primary challenges related to collecting the necessary data set at the computing center where the scientist has an allocation of sufficient scale to conduct the analysis. Current challenges in obtaining model output will only be increased in the next round of global climate model intercomparison, CMIP6 (Meehl et. al. 2014), as multiple modeling groups produce data sets with horizontal resolution less than the 100 kilometers typical of the CMIP5 data due to increases in their access to high performance computing resources.

In this paper, we document all aspects of the data analysis workflow involved in using CMIP5 data to characterize extra-tropical cyclones (ETC). This case study is presented to demonstrate some of the challenges in performing large dataset climate modeling analyses that are enabled by the CMIP5 and the Earth System Grid Federation (ESGF). The perspective is from both ordinary climate model analysts (the secondary authors) and a network professional (the lead author). It is not our intent to criticize either of these two projects but rather to help guide the enhancements that will be necessary to more fully enable this class of analyses in the next generation of distributed climate model output datasets.

To explicitly track ETCs we required simulated 6-hourly instantaneous 850hPa wind and equivalent sea level pressure fields. We confined our analysis to the present day (historical) and one future (rcp8.5) scenario datasets in CMIP5. Analysis included all individual ensemble members from all available climate models. The Toolkit for Extreme Climate Analysis (TECA) framework (Prabhat et. al. 2012) was used to simultaneously track the ETCs from the entire dataset using 750,000 cores on the Argonne Leadership Computing Facility (ALCF) IBM BG/Q Mira platform (http://www.alcf.anl.gov/) at the Argonne National Laboratory. Approximately 56 terabytes of data were downloaded using the Earth System Grid Federation (ESGF) portal software (http://pcmdi9.llnl.gov/esgf-web-fe/) to the National Energy Research Supercomputing Center (NERSC) facility, at the Lawrence Berkeley National Laboratory (http://www.nersc.gov/) and the entire data set subsequently transferred to ALCF for analysis.

The ETC analysis project required two large data sets from the CMIP5 project (http://cmip-pcmdi.llnl.gov/cmip5/data_portal.html). The sea level pressure and wind data sets were obtained from the *psl, ua,* and *va* variables from all model realizations of the *historical* and *rcp85* CMIP5 experiments at 6-hour time resolution. Although we required winds at only 850hPa, all vertical levels (usually 3) had to be downloaded as the ESGF does not currently permit subsetting operations. For at least one model, surface pressure (*ps*) was also required in order to complete the vertical interpolation from model coordinates to 850hPa. The data were downloaded from the globally-distributed archive to NERSC using the NERSC data transfer nodes, a set of fast servers optimized for data transfer, and staged on the NERSC global scratch filesystem, a multi-petabyte storage system accessible by all major NERSC resources.

The CMIP5 data sets are made available to the scientific community by means of the ESGF data portal. The ESGF portal provides a common interface to a distributed set of data repositories, which make the CMIP5 data available for use by scientists. The portal consists of a web interface to a data catalog, which provides search features and the ability to generate scripts for downloading data that match the search criteria. The scripts then transfer the data files from the data servers that store the data files and serve them to remote users (these data servers are called "data nodes") to the storage of the system running the script. ESGF data nodes that serve CMIP5 data are located around the world, typically at the modeling centers that participated in the CMIP5 experiments – the modeling centers that generated a particular data set typically have the ESGF data nodes that serve that data to the scientific community. These modeling centers are located in many different countries, because there are scientific and modeling groups in many different countries that are part of the CMIP5 experiments. Because of the distributed nature of the ESGF, assembling a data set that includes all the CMIP5 models requires downloading data from many individual sites, often over intercontinental distances. Figure 1 and Figure 2 list the data servers we used (20 in all).

Several replicas of the ESGF portal service are available – we used the master portal, hosted by the Program for Climate Model Diagnosis and Intercomparison (PCMDI) at the Lawrence Livermore National Laboratory. In addition to the portal, PCMDI hosts replicas of many CMIP5 data sets, as do a few other large data centers – this provides a redundant copy to mitigate risk of data loss that also allows for improved access by scientists in the United States (e.g. for data sets produced by modeling centers on other continents where data transfer performance to the US is slow). However, many of the individual six hourly data sets used in this analysis had not been replicated to PCMDI, which required us to transfer them to NERSC from their primary sources. The ESGF is further described in (Cinquini et. al. 2012).

The different components of the workflow took different amounts of time. These are summarized below.
- Data staging to NERSC from globally-distributed ESGF data nodes: 3 months
- Data pre-processing at NERSC: 2 weeks
- Replication of raw data set from NERSC to ANL: 2 days
- Pattern detection using TECA (750,000 cores) at ANL: 1.5 hours
- Verification of results using TECA at ANL: 10 minutes

The difference in elapsed time is quite significant, and we believe this illustrates the potential for dramatically improved scientific use of the CMIP5 data archive if data transfer performance could be improved.

Section 2 describes the workflow of downloading and assembling the approximately 56 terabytes of data from the globally distributed CMIP5 data archive to the NERSC facility. Section 3 describes the errors encountered and workflow solutions to remedy them. Section 4 details the actual data transfer rates realized from each of the ESGF data nodes and the reasons for the wide disparity in those rates. Finally, in section 5 we discuss the implications for CMIP6 and make some concrete suggestions to increase the data transfer rates using contemporary technologies.

## 2. Data Transfer Workflow

The ESGF portal allows a user to discover data, and create scripts to fetch the data. The process for transferring a data set (e.g. all data files for one variable for all realizations of one model for one experiment) consists of two parts: creating a "wget script" using the ESGF interface, and then running the wget script to transfer the data. The text below describes our experience in working with the ESGF portal in the manner suggested by the ESGF portal documentation, which is to use the portal interface to create wget scripts which are then run by hand to transfer the data. Another method of downloading data, using the synchro_data tool from IPSL, addresses many of the workflow issues we identify
 (http://forge.ipsl.jussieu.fr/prodiguer/wiki/docs/synchro-data).    However, the synchro_data tool was unknown to us when we assembled our data set (and synchro_data is not, as of this writing, integrated into the ESGF portal and documentation as a recommended tool for use in transferring data).

We found that the process for transferring data from ESGF requires manual intervention by people who are sophisticated in the use of the underlying components of the ESGF portal or the portal software itself. For example, while the ESGF portal search interface allows the selection of variables to narrow the search, these do not have any effect. One must filter over the search results in the portal using a text string to isolate individual variables. We learned this after the first wget scripts began to fetch extra data, and asked for help in the online support forum. In addition, if the number of files returned by a query exceeds a given threshold, the resultant script will transfer an incomplete data set. In this case, one must use undocumented query options to increase the number of files that the query can return in order to get a complete wget script for the data set.

The process for creating the wget scripts is as follows:
1. Select the data in the browser (for example, Project: *CMIP5*, Realm: *atmos*, Experiment: *historical*, Time: *6hr*, and then a model and a data node from which to download the data).
2. Add the data to the "data cart" (analogous to a e-commerce shopping cart)
3. Enter the desired variable name in the search box
4. Filter the results by variable (use search box and "filter over text" function)
5. Click "wget all selected" to generate and save the wget script
6. Check the wget script to ensure that the wget script did not hit the file limit (the script will contain a warning if the file count threshold is reached)
7. If the threshold is reached, modify the query and re-generate the wget script

This was done for each variable (*psl, ua,* and *va,* and also for *ps* in some cases), for each model.

After generating the wget scripts, the scripts must be run to transfer the data. The wget scripts are bash shell scripts that use the open source *wget* tool to transfer the data using HTTP, the same protocol used by web browsers to fetch pages from web servers. Each script downloads a set of files sequentially, one file at a time. There are several components to the data transfer workflow: credential management, running the script itself, and error recovery.

The ESGF tracks data download statistics, and so requires credentials in order to access and download data. The credentials are easy to obtain (creating an OpenID account via the ESGF portal at PCMDI took about 5 minutes). The wget scripts use Grid certificates to grant download permission for each file (this is built into the data servers and the wget scripts). In order to transfer data, one must have a valid certificate, typically based on one's ESGF credential. Grid certificates are used, and a common case is to use the wget script to generate a temporary certificate based on the OpenID account created via the ESGF portal. The certificate has a lifetime of three days – this means that the typical workflow involves refreshing the ESGF certificates by means of a wget script, and then running as many wget scripts as necessary for three days. After three days, the certificate times out and must be refreshed again before data downloads can recommence.

The wget script itself is a *bash* shell script containing the logic to refresh the certificates as well as to download a set of data files. The script also contains file checksum information for each file in the data set. After each file is transferred, the script uses the md5sum utility to calculate an MD5 checksum of the file, and compares that checksum to the checksum for the file included in the wget script by the ESGF portal. If the checksums do not match, an error is reported and the script attempts to transfer the file again. The run time of the wget scripts varies widely (from minutes to weeks), depending on the amount of data that must be transferred and the data transfer performance of the remote data server (see Figure 2). At NERSC, we used the NERSC data transfer nodes to run the wget scripts – these are high performance servers capable of transferring data at a rate of 1TB/hour as described at the end of Section 4, so the data transfer performance was determined by the remote data server and the network conditions between NERSC and the remote data server.

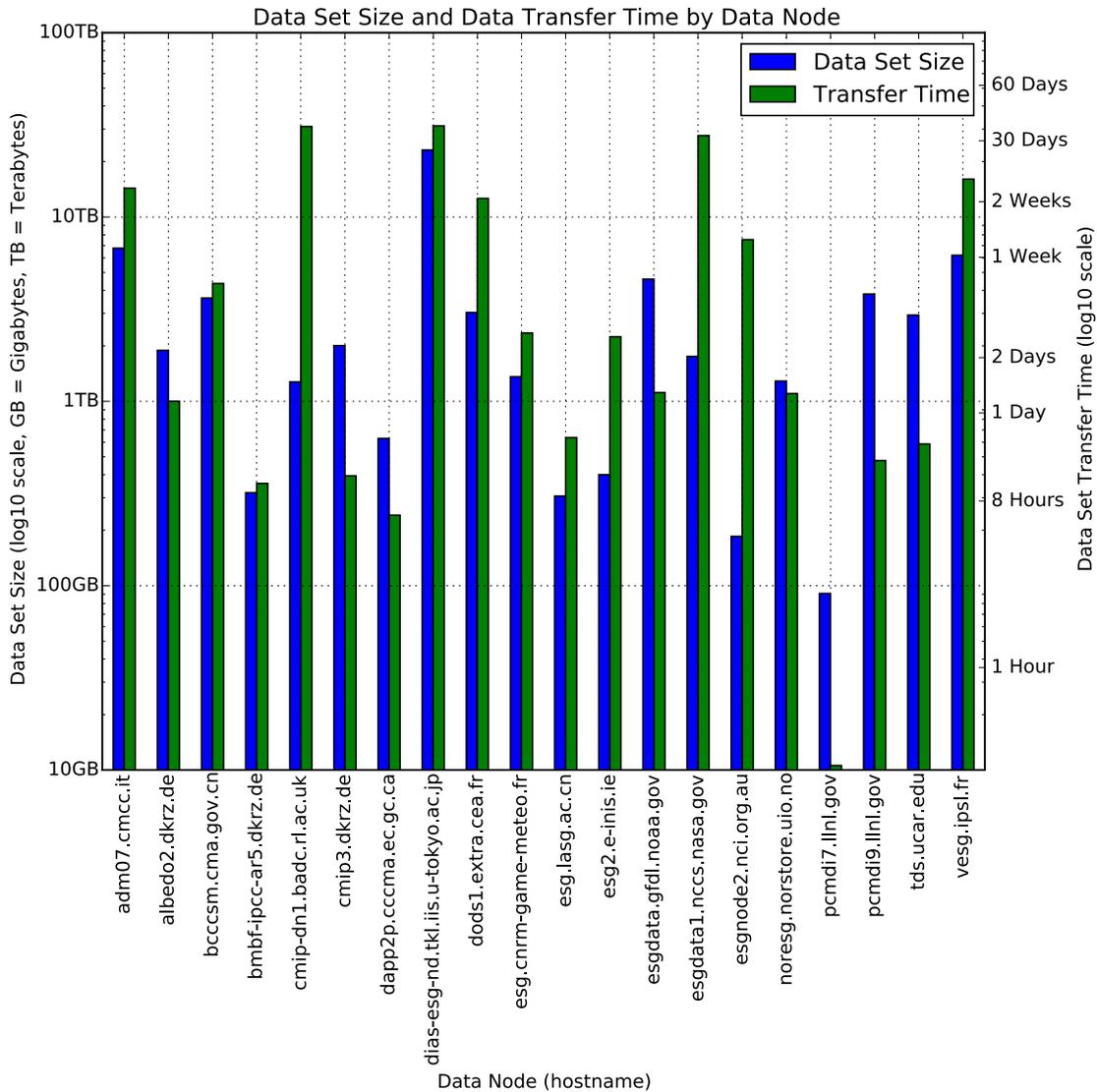

Figure 2: Data set size and single-threaded data transfer time for each ESGF data node. Note that human time to manage wget scripts, time spent checksumming the data, and other necessary functions are not included here – therefore, this represents the contribution of data transfer performance to the overall data staging time.

Error recovery takes multiple forms. We encountered several different error conditions, which are described in more detail in Section 3. However, the primary implication of the errors encountered and also of the necessity for credential management described above is that ***a real person has to actively manage the data transfer workflows***. In general, we checked on the data transfers at least once per day. In addition, because of the long run time of many of the scripts, it is unreasonable to assume that a stable login session to a data transfer node can be maintained over the run time of the script – system maintenance, travel, and myriad other events interrupted our sessions. We managed this by running each wget script in its own instance of the open source *screen* utility, which allowed us to start the data transfer running and then detach from the screen session (leaving the wget script running on the NERSC data transfer node), and reattach the screen session later via a different connection to the data transfer node to check on the progress of the script. Note also that because the credential lifetime is three days, and credential expiration causes all currently-running scripts to fail, credentials must be refreshed more often than every 72 hours. Because of scheduling realities, this often resulted in doing a credential refresh every two days, or logging in on the weekend in order to ensure credentials stayed fresh.

The wget scripts keep a status file for all the data they have transferred, and the status file is updated when a data file is successfully transferred and the checksum matches the value stored in the wget script. This is highly beneficial, because recovery from an error condition that causes the script to fail (e.g. credential expiration, storage quota exceeded, system maintenance) consists of simply running the same wget script again – the script picks up where it left off and resumes normal operation.

The data transfer workflow for each data set took the following form:
1. Ensure the credentials are fresh, and refresh them if they are stale (credential lifetime is 3 days)
2. Start a screen session labeled for the wget script we wish to run
3. Start a typescript session (using the unix *script* utility) within the screen session to capture the script output in case troubleshooting is necessary – this also collects performance statistics, since the wget scripts print this information as they run.
4. Run the wget script within the typescript session
5. Once a day, check for script completion, errors, and so on (but if the script might finish early, check more often)
6. Every three days, refresh the credentials
7. After the wget script completes, run the wget script again – if all data files were not successfully downloaded during the previous run, it will begin transferring the missing data files
8. Once the wget script has completed successfully (i.e. running the wget script again indicates that all files have already been successfully transferred), exit the typescript session
9. Exit the screen session
10. Mark the data set as done (manually touch a file named "DONE" in the data staging directory corresponding to the wget script)

Because of the amount of manual effort required to manage each wget script, there is a human limit to the number of these that can be run concurrently by a person. By the time the data staging was complete, we had accumulated and run 170 wget scripts. This is far beyond what a person can manage at one time, and shows a limitation of the present scripting model for managing large-scale data transfers.

The final data set for the project consisted of two large directory trees on a single filesystem – one tree containing the individual data sets for the models in the *historical* experiment and one for the *RCP 8.5* experiment. In addition to the data files themselves, numerous wget scripts, status files, and other workflow by-products were accumulated (though in terms of the aggregate data volume and file count the workflow management data is insignificant). The composition of the data sets, including the workflow management files, follows:

- *Historical* experiment
  - 28,000 files
  - 66 directories
  - 29444248373687 bytes (29.444 TB)
- *RCP 8.5* experiment
  - 29487 files
  - 98 directories
  - 26810787723285 bytes (26.81 TB)
- Total data volume: 56255036096972 bytes (56.255 TB)

## 3. Error Recovery and Data Transfer Workflow Management

We encountered several things that complicated the data staging workflow. A brief description of the most significant events follows.

It was not clear at the beginning how much data this project would require as the ESGF data portal does not give an estimate of data set size in advance of the actual transfer, so planning for filesystem space had to be done in an ad hoc fashion. We ran over NERSC filesystem quotas multiple times, and had to completely change filesystems once. Each of these events involved stopping all the data transfer workflows, and negotiating with the NERSC staff. To their credit, the NERSC consultants were very helpful and accommodating, but running out of storage space immediately stops all the workflows in an error condition and it takes human time to recover and get everything running again. If we had been able to determine the space required before staging the data, we could have set up an appropriate allocation *a priori* and would have avoided running over quota. While this might be considered a simple feature request, we believe information (such as the size of a data set) which makes it easier to manage large data sets is important to large-scale data workflows. In our view, the ESGF appears to have been designed to support downloading small subsets of large data archives, and in order to effectively support the large-scale analysis of the data in the CMIP5 archive (and in a future CMIP6 archive), several workflow improvements (such as this) will be necessary.

During our data staging work, one remote ESGF data node was taken down permanently, and its data sets were moved to another data node. In order to finish staging the data we had to create new wget scripts, manually incorporate the data we had into our directory structure for the new server, and then finish the workflow. The fact that the previous data node had been permanently replaced was not obvious at the start, so it took some time to troubleshoot.

Some data files did not match their checksums. In two cases, a file was simply published with the wrong checksum, and so the file was transferred repeatedly by the script until we discovered the condition. We then stopped the workflow and notified the data manager at the appropriate site, who investigated and corrected the error. In another case, a file failed its checksum after the first transfer, and passed the checksum after the second transfer – this behavior is transparent to the user, and the workflow was not interrupted. In a third case, two replicas of the data did not match, and we had to determine which site was serving the correct data using the checksums (we notified the other site, and the problem was corrected). These instances illustrate the value of the checksums beyond data integrity (in two of these cases the checksums alerted us to process problems or human error). However, the underlying causes of a majority of the checksum errors encountered point to processes that could be improved (e.g. replication, publication).

The performance of one site was so slow that the wget scripts for certain data sets were broken into smaller sub-scripts for performance reasons. In this case the per-download data rate was very slow, but each download would run at the same rate. This allowed us to scale up performance by running multiple scripts concurrently. However, this increased the number of scripts that had to be managed and increased the complexity by requiring the development of additional custom scripts.

## 4. Data Transfer Performance

Figure 1 shows the data transfer performance for the data files downloaded to NERSC from each of the ESGF data nodes. The left Y-axis shows data transfer performance on a log scale, from 100KB/sec to 1GB/sec. The right Y-axis shows the equivalent time required to download 1TB of data. The performance of the data transfers varied widely. Some data servers were very slow – the slowest had an average performance of 10 KB/sec. Fortunately the data from this particular server had been replicated at PCMDI, so the replica at PCMDI was used. Several data servers performed at between 100KB/sec and 200KB/sec, and there was no replica of their data at PCMDI – these servers were bottlenecks for the project. We do not believe the NERSC data transfer servers contributed materially to performance problems because of the high performance demonstrated when we transferred the data from NERSC to ALCF as described at the end of this section.

Figure 1: Data transfer rate performance statistics for ESGF data nodes plotted on a logarithmic scale.

Note that ten of the ESGF data nodes performed at 10MB/sec or less for a majority of the files downloaded from that data node. This is significantly less than the performance that is achieved by computing centers such as ALCF and NERSC. For comparison, 10MB/sec (80 Mbps) is roughly equivalent to the fastest US residential broadband speed characterized in the FCC's "Measuring Broadband America" 2014 online report ([http://www.fcc.gov/reports/measuring-broadband-america-2014#Figure3](http://www.fcc.gov/reports/measuring-broadband-america-2014#Figure3)) . Such transfer rates will transfer 1TB of data in about 28 hours and are achievable using networking technology (Fast Ethernet) from the mid-1990s. The fast servers and networks of 2014 are easily capable of 500MB/sec between major data centers when using modern tools, which corresponds to 1 TB in about 35 minutes. Much higher performance (greater than 1GB/sec) has been demonstrated in test environments (Rajendran et al. 2013). We believe that significant performance improvements could be made for the ESGF data nodes if the data centers which host ESGF data nodes made the appropriate resource commitments. These improvements would be twofold – better networking and data server infrastructure to improve per-connection performance, and better data transfer tools that make use of multiple simultaneous data transfer connections in parallel. These are further described in Section 5.

Low performance matters much less when the amount of data to be transferred is small. One could argue that there is no need for spending significant effort on improving performance for a server that hosts data sets that are a few gigabytes in size. However, the growth in data set size means that most major ESGF data nodes will be serving much larger data sets for CMIP6. In addition, as shown in fig. 2, it is already the case that our data transfers took a significant amount of time. The plot shows both the amount of data (left Y axis) and the time spent transferring the data (right Y axis) for each data node. Both fig. 1 and fig. 2 were obtained from the same data transfers.

The time shown in fig. 2 is only the amount of time spent transferring the data and writing the file to disk – it does not include the calculation of the file checksum for integrity verification after the transfer, and it does not include all the human time spent managing the workflow.

Even including only the active data transfer time, it is clear that staging many data sets takes a significant amount of time. The variation in performance level (as shown in fig. 1) accounts for part of the data transfer time, as does the volume of data that must be transferred from a particular data node. Many data nodes had total data transfer times of more than one day, and six had transfer times of more than two weeks. Specifically, node adm07.cmcc.it had a data transfer time of 1,433,705 seconds (16.5 days), cmip-dn1.badc.rl.ac.uk took 3,097,432 seconds (35.8 days), dias-esg-nd.tkl.iis.u-tokyo.ac.jp took 3,119,685 seconds (36.1 days, though the data volume was over 23TB), node dods1.extra.cea.fr 1,260,878 seconds (14.6 days), esgdata1.nccs.nasa.gov took 2,763,209 seconds (32 days), and vesg.ipsl.fr took 1,605,439 seconds (18.6 days). Because these times represent the total amount of time spent transferring data, and the data sets are composed of many files, a data transfer tool which transfers multiple files in parallel (such as Globus or synchro_data) could significantly reduce the amount of time required to transfer a data set. However, as we see in figure 1, some data nodes have significant room for improvement when the different data nodes are compared with each other.

Unless the data transfer servers (the ESGF data nodes in this case) have been tuned for high performance, the performance of the data transfer servers and tools is proportional to the distance between the data server and the server downloading the data. It is likely, for example, that for European computational facilities the European ESGF data servers would perform better because of the shorter distance of the data transfers. Also, the high performance achieved between PCMDI and NERSC might be due in part to this same phenomenon. However, because of the distributed and international nature of both climate science and the CMIP5 archive, it is reasonable to assume that any large-scale multi-model analysis of CMIP5 data will require long-distance data transfers no matter where the HPC facilities used by the analysts are located. Therefore, the ESGF data nodes, the networks they use, and the data transfer nodes at the computing centers must all be engineered to perform well even for long-distance (i.e. intercontinental) data transfers in order to support large-scale analysis of the data from CMIP5 and its successors. Note, however, that the high data transfer performance levels required to support the assembly of data sets on the scale tens to hundreds of terabytes need only be supported between the ESGF data centers and the major computing facilities which are capable of analyzing data sets of this scale. Because of this, we believe that such a performance improvement effort is practical and could be successful if the resources were committed.

Overall, between quota increase requests, wget script creation, workflow management, and failure recovery, the data transfer and staging portion of the project took about three months – from July 26th to October 20th 2013.

Later in the project, the aggregate raw data set was transferred from NERSC in California to the ALCF in Illinois across the ESnet, the Energy Sciences Network, (http://www.es.net/) using the Globus (https://www.globus.org/) fast transfer tool (Allen et al. 2012). Each of the two directory trees (one for the *historical* experiment and one for the *rcp85* experiment) were transferred using a single Globus transfer request each, running concurrently (two simultaneous requests in total). The entire 56TB data set was transferred from NERSC to ALCF in about 48 hours with a minimal commitment of human time. Excerpts from the output summaries from the Globus jobs are listed in the appendix. This performance difference is stark, and is an indication of what could be achieved through infrastructure investments at the major ESGF data centers. We are aware of some improvements that have been made (e.g. at ANU/NCI in Australia, BADC in the UK, and NASA in the USA) since October 2013. However, we believe much more can and should be done to enable high-performance data transfers from the CMIP5 archive data servers to major computational facilities worldwide where climate scientists perform large-scale multi-model data analyses.

## 5. Implications For Large Scale Data Analyses

It is clear from the difference in performance between the data transfer rates from NERSC to ALCF and the data staging to NERSC from the various ESGF data nodes that significant performance improvements for ESGF are within technical reach. However, in the absence of such improvements, data analysis projects that require a large amount of data from many ESGF data nodes (e.g. studies that include all CMIP5 models, or in the future all CMIP6 models) will be difficult to conduct because a dedicated data manager is essentially required for managing the data staging effort. We argue that the current circumstances significantly reduce the scientific value of the CMIP5 data set (and similar data sets, present and future) because of the difficulty of staging the data to conduct large-scale global analyses.

We note that many of the consumers of the CMIP5 data archive, especially in the climate change impacts community, require only small portions of the CMIP5 data sets to do their work. Data subsetting and remote analysis services are being developed to serve these needs and will reduce the time required to obtain the necessary data by reducing the volume of data. However, such services will not reduce the complexity of the access to the CMIP5 dataset as it is constructed presently.

Remote analysis services are also not likely to be useful for the type of global analysis we conducted. The tracking algorithm tool (TECA) requires significant CPU resources. It is unlikely that each and every CMIP5 data center will provide this level of computing capability for external users in the foreseeable future. Ours is one relatively straightforward usage of sub-daily data. The adoption of more sophisticated "Big Data" analysis techniques, particularly machine learning methods (Baldi et. al. 2014; Krizhevsky et. al. 2012; Monteleoni et. al. 2013), can exploit even larger, more multivariate portions of the CMIP dataset and could consume nearly as many compute cycles to analyze as was required to produce the model output. Hence, large data transfers to high performance computing centers will likely remain the only way to perform global multi-model analyses of high frequency CMIP data.

The protocols for the next version of the Couple Model Intercomparison Project (CMIP6) are being negotiated at the present time (Meehl et. al. 2014). Regardless of which subprojects are adopted, the total dataset size for CMIP6 (as well as the size of data sets from individual model simulations) is virtually certain to be substantially larger than for CMIP5 due to advances in both model development as well as high performance computing technologies. In particular as the models' spatial resolutions become finer, the high-resolution output becomes more informative and demand for such data will likely increase. We recommend that, as part of the CMIP6 effort, significant performance improvements be made to the data transfer infrastructure that supports the distribution of climate data. We expect that both software improvements (e.g. better workflow tools, and the adoption of high-performance data transfer protocols as a replacement for single-threaded HTTP) and systems improvements (ensuring that high-performance systems and networks are deployed and maintained for serving ESGF data sets) will be required. The authors believe several higher-level changes to the ESGF would substantially improve performance:
1) Major climate data centers should adopt current best practices such as the Science DMZ model (Dart et al. 2013) for high-performance data transfer infrastructure, as many other major computing centers (including ALCF and NERSC) do. If the major climate data centers dedicated a portion of their infrastructure specifically to the task of high-performance data transfer to remote facilities as outlined in the Science DMZ model, large-scale data transfers to major scientific computing centers could be made routine.
2) Globus (or an equivalent high-performance tool) file transfers should be enabled for all files for reasons of both usability and performance. Currently a subset of the CMIP5 data set, mostly monthly fields, can be obtained via Globus. This is useful for providing data with low time resolution to a large number of individual analysis teams, and makes good use of the better usability of the Globus tools. However, the high performance parallel transfer methods used by Globus are far better suited to the large-scale data transfers required for staging high-resolution data for analysis than the HTTP protocol which is the current standard in ESGF. While Globus is not the only high-performance data transfer tool available, the current use of Globus in the ESGF and the support by Globus of the authentication mechanisms used by the ESGF make Globus a logical choice for consideration as the standard high-performance data transfer tool in the ESGF. High performance tools are required for efficiently assembling the large-scale data sets necessary to conduct analyses like the ETC analysis we describe in section 1. The synchro_data tool partially addresses some of these issues (e.g. it runs multiple parallel wget processes at a time), but synchro_data is still bound to wget as an underlying data transfer mechanism, and so synchro_data can only go as fast as wget can go.
3) The ESGF should explicitly support large-scale data analysis at major computational facilities. In addition to data transfer performance, this requires workflow improvements to support large data staging efforts. These include providing advance estimates of requested data volume to allow users to plan for storage requirements, and a reworking of the certificates used to authenticate users (e.g. extending certificate lifetime for large data transfers). Additionally, the individual modeling centers should carefully consider whether certificates are necessary to protect their data, as this additional complication has proven difficult for many analysts.
4) Sufficient financial resources must be provided to the ESGF data centers to support network professionals to tune their systems and networks for intercontinental data transfers. Data managers at the centers must also be adequately supported to maintain quality control.
5) Finally, data transfer performance metrics that provide realistic expectations of data transfer performance between ESGF data nodes and major computing centers should be regularly collected and published. This would allow analysts to realistically assess the feasibility of analyses with large-scale data requirements. In addition, the open publication of performance metrics would allow the scientific community to assess the practical contributions of individual data centers to large-scale data analysis efforts.

We conclude by highlighting the World Climate Research Program survey conducted after the 5th IPCC Assessment Report to determine what went well and what did not for analyses of the CMIP5 simulations. CMIP5 data users have indicated a need for improving the data transfer infrastructure used in climate science (Eyring and Stouffer 2013). The finding for section 10, "Data search and support" is:

> *Overall: Improve logistics of CMIP6, as a software, analysis and management problem. These things - how to get the data, how to analyze it, how to manage it - were fundamentally limiting in CMIP5 for many groups.*

Clearly, there is a need to focus resources on improving data transfer performance between the ESGF data nodes and major computing centers. It is also clear that significant improvements are within reach using current technologies. All that is required is for the resources to be committed.

References:

Allen, B., J. Bresnahan, L. Childers, I. Foster, G. Kandaswamy, R. Kettimuthu, J. Kordas, M. Link, S. Martin, K. Pickett, S. Tuecke 2012: Software as a Service for Data Scientists. Comm. Assoc. Computing Machinery 55, 81-88.

Baldi, P., P. Sadowski, D. Whiteson 2014: Searching for Exotic Particles in High-energy Physics with Deep Learning. Nature Comm. 5 doi:10.1038/ncomms5308

Cinquini, L., Jet Propulsion Lab. (JPL), California Inst. of Technol., Pasadena, CA, USA; Crichton, D. ; Mattmann, C. ; Bell, G.M. ; Drach, B. ; Williams, D. ; Harney, J. ; Shipman, G. ; Feiyi Wang ; Kershaw, P. ; Pascoe, S. ; Ananthakrishnan, R. ; Miller, N. ; Gonzalez, E. ; Denvil, S. ; Morgan, M. ; Fiore, S. ; Pobre, Z. ; Schweitzer, R. 2012: The Earth System Grid Federation: An open infrastructure for access to distributed geospatial data. 2012 IEEE 8th International Conference on E-Science (e-Science), doi: 10.1109/eScience.2012.6404471

Dart, E., L. Rotman, B. Tierney, M. Hester, J. Zurawski 2013: The Science DMZ: A Network Design Pattern for Data-Intensive Science, Proceedings of SC13: The International Conference for High Performance Computing, Networking, Storage and Analysis, Denver CO, USA. November 17-21, 2013 doi:10.1145/2503210.2503245

Eyring, V. and R. Stouffer 2013: Synthesis WCRP Coupled Model Intercomparison Project Phase 5 (CMIP5) Survey, Presentation to the 17[th] session of the Working Group on Climate Modeling, Victoria Canada October 1-3, 2013
http://www.wcrp-climate.org/wgcm/references/SynthesisCMIP5Survey_131001.pdf

Kharin, V.V., F. W. Zwiers, X. Zhang, M. Wehner 2013: Changes in temperature and precipitation extremes in the CMIP5 ensemble, Clim. Change 119, 345-357 doi:10.1007/s10584-013-0705-8.

Krizhevsky, A., I. Sutskever, G. Hinton, 2012: ImageNet Classification with Deep Convolutional Neural Networks, in Advances in Neural Information Processing Systems 25, F. Pereira, C.J.C. Burges, L. Bottuu, K.Q. Weinberger, eds.

Meehl, G. A., R. Moss, K. E. Taylor, V. Eyring, R. J. Stouffer, S. Bony, and B. Stevens 2014: Climate Model Intercomparison: Preparing for the Next Phase, Eos, Trans. AGU 95, 77

Monteleoni, C., G. A. Schmidt, S. McQuade 2013: Climate Informatics: Accelerating Discovering in Climate Science with Machine Learning, Computing in Science and Engineering 15, 32-40

Prabhat, O. Ruebel, S. Byna, K. Wu, F. Li, M. Wehner and W. Bethel (2012) TECA: A Parallel Toolkit for Extreme Climate Analysis, International Conference on Computational Science, ICCS 2012, Workshop on Data Mining in Earth System Science, Procedia Computer Science 9, 866-876

Rajendran, A., P. Mhashilkar, H. Kim, D. Dykstra, G. Garzoglio, I. Raicu 2013: Optimizing Large Data Transfers over 100Gbps Wide Area Networks, Proceedings of IEEE/ACM International Symposium on Cluster, Cloud and Grid Computing (CCGrid), http://datasys.cs.iit.edu/publications/2013_CCGrid13-100Gbps.pdf

Sillmann, J., V.V. Kharin, X. Zhang, F.W. Zwiers and D. Bronaugh, 2013: Climate extreme indices in the CMIP5 multi-model ensemble. Part 1: Model evaluation in the present climate. J. Geophys. Res.: Atmospheres, 118, 1-18, doi:10.1002/jgrd.50203

Taylor, K.E., R. J. Stouffer, and G. A. Meehl, 2012: An Overview of CMIP5 and the Experiment Design. Bull. Amer. Meteor. Soc., 93, 485–498. doi: http://dx.doi.org/10.1175/BAMS-D-11-00094.1

Michael F. Wehner, Kevin Reed, Fuyu Li, Prabhat, Julio Bacmeister, Cheng-Ta Chen, Chris Paciorek, Peter Gleckler, Ken Sperber, William D. Collins, Andrew Gettelman, Christiane Jablonowski (2014) The effect of horizontal resolution on simulation quality in the Community Atmospheric Model, CAM5.1. *Journal of Modeling the Earth System* 06, 980-997. doi:10.1002/2013MS000276

## Acknowledgements

This work was supported by the Regional and Global Climate Modeling Program of the Office of Biological and Environmental Research in the U.S. Department of Energy Office of Science under contract number DE-AC02-05CH11231. ESnet is funded by the U.S. Department of Energy, Office of Science, Office of Advanced Scientific Computing Research (ASCR). ESnet is operated by Lawrence Berkeley National Laboratory, which is operated by the University of California for the U.S. Department of Energy under contract DE-AC02-05CH11231. Calculations were performed at the National Energy Research Supercomputing Center (NERSC) at the Lawrence Berkeley National Laboratory. NERSC is a DOE Office of Science User Facility supported by the Office of Science of the U.S. Department of Energy under Contract No. DE-AC02-05CH11231. This research used resources of the Argonne Leadership Computing Facility, which is a DOE Office of Science User Facility supported under contract DE-AC02-06CH11357. The authors are grateful for help and support from the NERSC consultants who provided a generous temporary filesystem quota increase in support of this effort. The authors are also grateful for a grant of time on the Mira system at the ALCF from the Director's Reserve. The authors wish to thank Venkatram Vishwanath of the ALCF for his collaboration and assistance with I/O performance tuning on the Mira system at the ALCF. In addition, the ESGF community support forums were a valuable resource, both for interacting with data managers at ESGF sites and for soliciting help when problems were encountered.

## Appendix 1 – Globus Transfer Job Summaries

This section contains an excerpt from the summary output from the Globus system for the data transfers of the *historical* and *RCP8.5* data sets from NERSC to ALCF. These transfers ran concurrently, and the elapsed time includes the time spent verifying the integrity of the data after transfer using checksums.

| | |
|---|---|
| Task ID: | dc40346a-4d5d-11e3-9a00-12313d2005b7 |
| Request Time: | 2013-11-14 18:52:07Z |
| Completion Time: | 2013-11-16 18:28:31Z |
| Total Tasks: | 28067 |
| Source Endpoint: | nersc#dtn |
| Destination Endpoint: | alcf#dtn_mira |
| Files: | 28000 |
| Directories: | 66 |
| Bytes Transferred: | 2.94442E+13 |
| MBits/sec: | 1374.43 |
| Faults: | 0 |

Table A1: Globus transfer summary excerpt for historical data set.

| | |
|---|---|
| Task ID: | e06d9578-4d5d-11e3-9a00-12313d2005b7 |
| Request Time: | 2013-11-14 18:52:14Z |
| Completion Time: | 2013-11-16 15:42:37Z |
| Total Tasks: | 29586 |
| Source Endpoint: | nersc#dtn |
| Destination Endpoint: | alcf#dtn_mira |
| Files: | 29487 |
| Directories: | 98 |
| Bytes Transferred: | 2.68108E+13 |
| MBits/sec: | 1328.722 |
| Faults: | 0 |

Table A2: Globus transfer summary excerpt for RCP8.5 data set

# Appendix 2: A comment on the final status of this paper

This paper was originally submitted to the Journal of Advances in Modeling the Earth System. We felt that journal was an appropriate venue as climate model data distribution to the scientific community is a critical piece in modeling the earth system. The version presented here was the 2nd and final attempt to satisfy the review comments. We maintain that despite the difficulties in setting up the extraction of such large datasets from the CMIP5 archive, the limiting factor remains network performance. The ESNET network serving the NERSC center is among the fastest in the world. Nonetheless, as figures 1 and 2 illustrate, network performance in this exercise was unacceptably slow. We traced this to two factors. The first was a lack of gridftp access to the datasets we needed despite a GLOBUS capability in the ESGF software. The second was that networks were not appropriately tuned to achieve high bandwidths.

Reviewers were not convinced of this conclusion and asked us to redo some transfers with undocumented systems (at least within the ESGF webpages). These were the synchro_data tool (http://forge.ipsl.jussieu.fr/prodiguer/wiki/docs/synchro-data) and esgf-pyclient (http://esgf-pyclient.readthedocs.org/en/latest/index.html).

We were more than willing to do this, in order to settle the question. However, we were stymied by a illegal hack of some of the ESGF data nodes. The resultant shutdown of the entire ESGF system during most of 2015 and the absence of any clear indication as to its resurrection prevented us from any further testing. After several patient responses from the chief editor while we waited for ESGF functionality to return, a new editor asked us to withdraw the paper because of this unavoidable delay. He also indicated that under his leadership, this type of study was not in the direction he wanted the journal to go and suggested resubmission elsewhere.

At this time (August 2017), the ESGF and the CMIP5 archive has largely (but not completely) been restored to functionality. Also, neither the synchro_data nor esgf-pyclient tools are publicized by the cmip5 webpages and remain largely unknown within the climate model analysis community. As part of the recovery from the hack, the collection of ESGF data servers is smaller with certain datasets replicated at other locations than they were during our data transfer exercise. Hence, some of the results reported here are out of date. However, at this time, the interests and priorities of the authors have shifted and we have elected to leave further investigation of large CMIP5/6 data transfers to others. As the datasets required for this study are not GLOBUS enabled, we expect that a repeat of this exercise would still not yield acceptable data transfer rates. We maintain that this is critical for the success of CMIP6. For as models attain higher horizontal resolution, the high frequency data is far more interesting, not just for our intended storm tracking study but for other classes of analyses.